\documentclass[11pt]{article}
\usepackage{amsmath,amssymb,color,graphics,epsfig,cite}
\usepackage{amsfonts}
\newcommand{\be}{\begin{equation}}
\newcommand{\ee}{\end{equation}}
\newcommand{\bea}{\setlength\arraycolsep{2pt} \begin{eqnarray}}
\newcommand{\eea}{\end{eqnarray}}

\def\ft#1#2{{\textstyle{\frac{\scriptstyle #1}{\scriptstyle #2} } }}
\def\fft#1#2{{\frac{#1}{#2}}}

\def\0{{\sst{(0)}}}
\def\1{{\sst{(1)}}}
\def\2{{\sst{(2)}}}
\def\3{{\sst{(3)}}}
\def\4{{\sst{(4)}}}
\def\5{{\sst{(5)}}}
\def\6{{\sst{(6)}}}
\def\7{{\sst{(7)}}}
\def\8{{\sst{(8)}}}
\def\sst#1{{\scriptscriptstyle #1}}

\begin{document}

\begin{center}
{\large {\bf Global Distinctions Between New Electrovacuum and Kundt Class}}

\vspace{20pt}

Liang Ma and H. L\"u\footnote{maliang0@tju.edu.cn\ \ \ mrhonglu@gmail.com}

\vspace{10pt}

{\it Center for Joint Quantum Studies and Department of Physics,\\
School of Science, Tianjin University, Tianjin 300350, China }

\vspace{40pt}

\underline{ABSTRACT}
\end{center}

It was observed \cite{Ovcharenko:2026uxi} that our recently constructed electrovacuum \cite{Ma:2026ima}, supported by electromagnetic fields, can be locally transformed to a special Petrov type D solution within the general Kundt class, which is not an electrovacuum but instead describes a spacetime generated by (accelerating) electric and magnetic charges. In this short note, which serves as a supplemental to \cite{Ma:2026ima}, we analyse the major global differences between the two locally-equivalent solutions.

\vspace{40pt}

%\vfill {\footnotesize mrhonglu@gmail.com}

%{\footnotesize \hoch{*}Corresponding author}

%\thispagestyle{empty}
%\pagebreak
%\voffset=-40pt
%\setcounter{page}{1}

%\tableofcontents
%\addtocontents{toc}{\protect\setcounter{tocdepth}{2}}

%\newpage

\section{The global distinctions}

General Petrov type D solutions of Kundt class, describing the spacetimes generated by (accelerating) electric and magnetic charges were presented in \cite{Ovcharenko:2026uxi}. For our purposes, we consider only the particular solution relevant to our discussion, given by
\bea
ds^2&=&p^2\Big(-Qd\tau^2+\frac{dq^2}{Q}\Big)+\frac{P}{p^2}dy^2+\frac{p^2}{P}dp^2\,,\cr
%%%
A_{\1}&=& q_e q\, d\tau-\frac{q_g }{p}dy\,,\qquad Q=\epsilon_0-\epsilon_2q^2\,,\qquad P=\epsilon_2 p^2-(q_e^2+q_g^2)\,.\label{kundt}
\eea
For simplicity, we set the magnetic charge $q_g=0$. The preservation of the spacetime signature requires $\epsilon_2>0$. Without loss of generality, we choose $\epsilon_2=1$. The coordinate $p$ is noncompact, running from $p=q_e$ to infinity. The degeneracy of the Killing vector $\partial_y$ at $p=q_e$ implies that it must be periodic with charge-dependent periodicity $\Delta y=2\pi q_e$. It is straightforward to verify that the spacetime carries non-vanishing electric charges (more precisely electric flux), given by
\be
Q_e=\frac{1}{4\pi}\oint*F_{\2}=\frac{1}{4\pi} \oint dy\,\int_{q_e}^\infty dp\,\frac{q_e}{p^2}=\ft12 q_e\,.
\ee
Thus, the Kundt solution should not be regarded as an electrovacuum since the spacetime contains charged matter. The Maxwell field can be turned off by setting $q_e=0$, and the metric reduces flat, given by
\be
ds^2=dp^2 + p^2\Big(-Qd\tau^2+\frac{dq^2}{Q}\Big) + dy^2\,.
\ee
It is a direct product of $\mathbb R$ and a Ricci-flat cone of dS$_2$ (de Sitter). The resulting flat metric is not the globally-defined Minkowski spacetime, without further analytical continuation. A serious pathology of the $q_e\rightarrow 0$ limit is that we must impose a topology change of $y$ by hand, from a periodic circle with $\Delta y=2\pi q_e\rightarrow 0$ to a real line, in order to avoid singularity. This indicates the Kundt solution does not have a smooth flat limit, since $q_e\rightarrow 0$ does not reproduce the $q_e=0$ solution.

We now perform the coordinate transformation prescribed in \cite{Ovcharenko:2026uxi}
\bea
\tau=\frac{\mathcal{E}}{\sqrt{\epsilon_0}}t\,,\quad p=\frac{1}{\mathcal{E}\sqrt{1-\mathcal{E}^2r^2(1-x^2)}}\,,\quad q=\frac{\mathcal{E}\sqrt{\epsilon_0}rx}{\sqrt{1-\mathcal{E}^2r^2(1-x^2)}}\,,\quad y=\mathcal{E}^{-1}\phi\,,
\eea
and set $q_e = -\mathcal{E}^{-1}$. This brings us to the electrovacuum solution \cite{Ma:2026ima}
\bea
ds^2&=&\frac{1}{\big(1- \mathcal{E}^2 r^2
   (1-x^2)\big)^2}\Big[-(1-\mathcal{E}^2 r^2) dt^2+\frac{dr^2}{1-\mathcal{E}^2 r^2}+r^2\frac{ dx^2}{1-x^2}\Big]+r^2 (1-x^2) d\phi^2\,,\cr
   %%%
A_{\1}^{\mathrm{ele}}&=&-\frac{\mathcal{E}r x }{\sqrt{1-\mathcal{E}^2r^2 (1-x^2) }}dt\,.\label{vacuum}
\eea
The fact that $({*F})_{x\phi}$ is an odd function of $x\in [-1,1]$ implies that the spacetime is uncharged everywhere, and hence a vacuum. Furthermore, setting ${\cal E}=0$ turns off the Maxwell field and yields the globally well-defined Minkowski spacetime in spherical polar coordinates. Note that the local transformation requires a peculiar relation $q_e {\cal E}=-1$, which implies $(q_e,{\cal E})$ cannot be simultaneously zero; however, it is consistent with that the turning off of the Maxwell field can lead to globally different flat metrics.

\section{The BR analogy}

At first sight, it may seem surprising that locally-related solutions can be so globally different. Even the conserved charges can differ between such solutions. However, this is not uncommon in general relativity and we present the Bertotti-Robinson (BR) electrovacuum \cite{Bertotti:1959pf,Robinson:1959ev} as a closely related example. The BR electrovacuum is given by
\bea
ds^2&=&\frac{1}{1+B^2r^2(1-x^2)}\Big[
-(1+B^2r^2)dt^2+\frac{dr^2}{1+B^2r^2}+r^2\big(\frac{dx^2}{1-x^2}+(1-x^2)d\phi^2\big)
\Big]\,,\cr
%%%
A_{\1}&=&\Big(\frac{1}{B\sqrt{1+B^2r^2(1-x^2)}}-\fft{1}{B}\Big)d\phi\,.\label{br}
\eea
This spacetime is supported by a magnetic field, but is free of magnetic charges. The metric reduces to the globally well-defined Minkowsk spacetime by simply setting $B=0$, in which case the Maxwell field vanishes. This property is shared by our electrovacuum \eqref{vacuum}.

By performing the coordinate transformation
\be
R^2 = \fft{r^2 x^2}{1+B^2 r^2(1-x^2)}\,,\qquad X^2 = \fft{1}{1+B^2 r^2(1-x^2)}\,,
\ee
from \(\{r, x\}\) to \(\{R, X\}\), the BR solution \eqref{br} can be locally recast into the manifest \(\mathrm{AdS}_2 \times S^2\) form, given by
\be
ds^2=-(1+B^2R^2)dt^2+\frac{dR^2}{1+B^2R^2}+
B^{-2}\Big(\frac{dX^2}{1-X^2}+(1-X^2)d\phi^2\Big)\,,\quad A_{\1}=B^{-1} X\,d\phi\,.\label{ads2s2}
\ee
This AdS$_2\times S^2$ solution carries non-zero magnetic charge (more precisely magnetic flux), similar to the Kundt solution \eqref{kundt}
\bea
Q_m=\frac{1}{4\pi}\oint_{S^2}F_{\2}=B^{-1}\,.
\eea
Indeed, this $Q_m B=1$ relation is analogous to our earlier $q_e {\cal E}=-1$ relation. To turn off the Maxwell field, we cannot send $B\rightarrow \infty$. Instead, we rescale the compact coordinates as $X\rightarrow B X $ and $\phi\rightarrow B Y$, and then send $B\rightarrow 0$. In this limit, the Maxwell field vanishes and the metric becomes flat as $ds^2 = -dt^2 + dR^2 + dX^2 + dY^2$. The limit is singular at the charge sector, but smooth at the metric level since $\Delta Y= 2\pi/B\rightarrow \infty$.  This should be contrasted with the Kundt solution which has no smooth flat limit as discussed.

To conclude, the BR electrovacuum and AdS$_2\times S^2$ are globally distinct, just as new electrovacuum \eqref{vacuum} and the local Kundt solution \eqref{kundt} are globally distinct. The electrovacuum \eqref{vacuum} describes an uncharged spacetime supported by external electromagnetic fields, whilst the Kundt solutions are not electrovacua, but spacetimes generated by (accelerating) electric or magnetic charges. There are currently two well-known electrovacua Bonnor-Melvin and BR solutions, and ours provides a new one.

\section*{Acknowledgement}

We are grateful to James Liu and Hai Huang for useful discussions. L.M.~is supported in part by National Natural Science Foundation of China (NSFC) grant No.~12447138, Postdoctoral Fellowship Program of CPSF Grant No.~GZC20241211, the China Postdoctoral Science Foundation under Grant No.~2024M762338 and the National Key Research and Development Program No.~2022YFE0134300. H.L.~is supported in part by the NSFC grants No.~12375052 and No.~11935009. Both are also supported in part by the Tianjin University Self-Innovation Fund Extreme Basic Research Project Grant No.~2025XJ21-0007.

\appendix

\end{document}